\documentclass[sigconf,final]{acmart}
\usepackage{booktabs} 
\usepackage[utf8]{inputenc}
\usepackage[english]{babel}
\usepackage{multicol}
\usepackage{graphicx}
\usepackage{subcaption}
\usepackage{multirow}
\usepackage{tcolorbox}
\usepackage{xcolor}
\usepackage{colortbl,hhline}
\usepackage[export]{adjustbox}
\usepackage{adjustbox}
\usepackage{comment}
\usepackage{paralist}
\usepackage[textsize=scriptsize]{todonotes}
 
\copyrightyear{2020}
\acmYear{2020}
\setcopyright{acmcopyright}\acmConference[JCDL '20]{Proceedings of the ACM/IEEE Joint Conference on Digital Libraries in 2020}{August 1--5, 2020}{Virtual Event, China}
\acmBooktitle{Proceedings of the ACM/IEEE Joint Conference on Digital Libraries in 2020 (JCDL '20), August 1--5, 2020, Virtual Event, China}
\acmPrice{15.00}
\acmDOI{10.1145/3383583.3398527}
\acmISBN{978-1-4503-7585-6/20/06}

\begin{document}
\fancyhead{}
\title{Characterising authors on the extent of their paper acceptance: A case study of the Journal of High Energy Physics}

\author{Rima Hazra}
\affiliation{%
  \institution{Indian Institute of Technology}
  \city{Kharagpur}
  \country{India}
  \postcode{P.O -- 721302}
}
\email{to_rima@iitkgp.ac.in}

\author{Aryan}
\affiliation{%
  \institution{Indian Institute of Technology}
  \city{Kharagpur}
  \country{India}
  \postcode{P.O -- 721302}
}
\email{aryankgp1576@gmail.com}

\author{Hardik Aggarwal}
\affiliation{%
  \institution{Indian Institute of Technology}
  \city{Kharagpur}
  \country{India}
  \postcode{P.O -- 721302}
}
\email{hardik8464@gmail.com}

\author{Matteo Marsili}
\affiliation{%
  \institution{ICTP}
  \city{Trieste}
  \country{Italy}
  \postcode{P.O -- 34151}
}
\email{marsili@ictp.it}

\author{Animesh Mukherjee}
\affiliation{%
 \institution{Indian Institute of Technology}
 \city{Kharagpur}
 \country{India}
 \postcode{P.O -- 721302}
 }
 \email{animeshm@cse.iitkgp.ac.in}

\renewcommand{\shortauthors}{Hazra et al.}

\begin{abstract}
New researchers are usually very curious about the recipe that could accelerate the chances of their paper getting accepted in a reputed forum (journal/conference). In search of such a recipe, we investigate the profile and peer review text of authors whose papers almost always get accepted at a venue (Journal of High Energy Physics in our current work). We find authors with high acceptance rate are likely to have a high number of citations, high $h$-index, higher number of collaborators etc. We notice that they receive relatively lengthy and positive reviews for their papers. In addition, we also construct three networks -- co-reviewer, co-citation and collaboration network and study the network-centric features and intra- and inter-category edge interactions. We find that the authors with high acceptance rate are more `central' in these networks; the volume of intra- and inter-category interactions are also drastically different for the authors with high acceptance rate compared to the other authors. Finally, using the above set of features, we train standard machine learning models (random forest, XGBoost) and obtain very high class wise precision and recall. In a followup discussion we also narrate how apart from the author characteristics, the peer-review system might itself have a role in propelling the distinction among the different categories which could lead to potential discrimination and unfairness and calls for further investigation by the system admins.
\end{abstract}

%
%


\maketitle

\section{Introduction}
\label{sec:intro}
\begin{figure}[!tbh]
\centering
\setlength\tabcolsep{0pt}
\renewcommand{\arraystretch}{0.8}%
\begin{tabular}{@{}c@{}c@{}c@{}c@{}}
  \includegraphics[width=0.65\hsize,height=5.5cm]{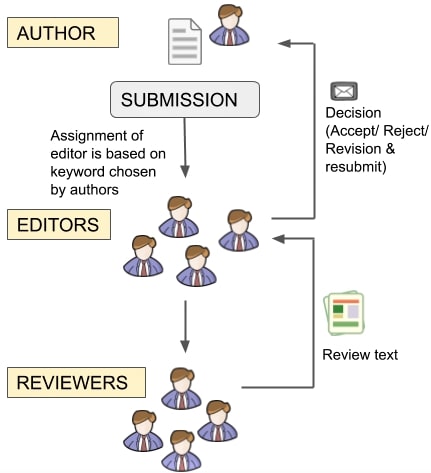} \\
\end{tabular}
\caption{The JHEP peer review workflow.}  
\label{fig:peer_review_system}
\end{figure}

Publishing new research in journals/conferences is a common practice in the scientific community. It is noticed that papers of few authors consistently get accepted in journals whereas papers of certain other authors get rarely accepted\footnote{https://www.sciencemag.org/careers/2018/12/yes-it-getting-harder-publish-prestigious-journals-if-you-haven-t-already}. An intriguing question thus is what makes the papers of certain authors almost always eligible for acceptance. Is there a special recipe that they follow in preparing their manuscripts? Does it depend on their position in the collaboration/citation network? Does their experience or their $h$-index matter? Does the diversity in the topics that they work on help escalate the acceptance? The present paper attempts to delve into some of these questions and characterise authors based on their paper acceptance profile. We base our investigations on a dataset obtained from the Journal of High Energy Physics that has information about authors, papers written by them, citations obtained by them and the review reports written by expert referees for each of their accepted paper. The overall peer review workflow for this journal is illustrated in Figure~\ref{fig:peer_review_system}. In a nutsell the workflow is as follows -- once an author submits a paper, the system allocates the submission to an editor based on a simple keyword matching technique. The editor then handles the paper and chooses one or more competent referees who are experts in the area and can judge the technical merit of the paper. The referee(s) in turn read the paper and send their review report(s) to the editor. The editor reads the review(s) and takes a decision to either accept, reject or invite the authors to revise and resubmit. The revise and resubmit decision re-instantiates the same workflow described above once again and the cycle continues until the paper is eventually accepted or rejected. 

We categorize the authors in this dataset into three classes based on the fraction of their papers accepted to the journal. We calculate the \textit{acceptance rate} ($ACC$) of an author as the ratio of the number of papers accepted to the number of papers submitted by the author to the journal. For each of the three categories (discussed below) we analyse a bunch of interesting features that are drawn from the collaboration/citation network of an author as well as the peer reviews received by the different accepted papers of the authors. We find that these features are considerably different across the three $ACC$ classes.

\vspace{-0.2cm}
\subsection{Our contributions}
 We categorize the authors into three classes based on their acceptance rate. Authors whose papers are consistently accepted for publication and have high $ACC$ are placed in the class $ACC_{high}$; authors whose papers are rarely accepted and have low $ACC$ are placed in the class $ACC_{low}$ and authors who are neither in $ACC_{high}$, nor in $ACC_{low}$ and have moderate $ACC$ are placed in $ACC_{mid}$. We explain the process of author categorization in details in section~\ref{sec:categorization}.
 Our main contributions are threefold.
\begin{enumerate}
    \item Rigorous analysis of the profile and peer review based features of authors belonging to each category. 
    \item Analyzing inter-category and intra-category interaction and network-centric properties obtained from three different networks -- (i) the co-reviewer network ($CRN$), (ii) the collaboration network ($CON$) and (iii) the co-citation network ($CCN$). 
    \item Early prediction of an author's category based on the profile, peer review data and network-centric features.
\end{enumerate}
Toward the first objective, we extract various features representing an author. These features are divided into two types -- (i) author's profile based features ($AP_f$) and (ii) features based on peer review data ($PE_f$). Author's profile based features ($AP_f$) comprises citation count ($C_{cnt}$), topic diversity ($T_{div}$), experience ($E_{cnt}$) and $h$-index~\cite{Hirsch:2005} ($H_{ind}$). Peer review based features ($PE_f$) consists of sentiment of review text ($SNT_{r}$), length of the review text ($L_r$), reviewer diversity ($R_{div}$) and editor diversity ($Ed_{div}$).

In addition, we extract various features -- centrality values, clustering coefficient, core-periphery structure etc. from the three different type of networks mentioned above. These networks are defined below.

\noindent \textbf{(i) Co-reviewer network ($CRN$)}: In this article, we introduce a co-reviewer network. Each author is considered as a node in the network and two authors are connected by an edge if their papers are reviewed by the same reviewer. In addition, we also prepare an induced co-reviewer graph for the three different author categories.\\
\noindent \textbf{(ii) Collaboration network ($CON$)}: Each author in this network is considered as a node and two authors are connected by an edge if they co-authored in a paper. We also prepare the induced collaboration networks of the authors of each category.\\
\noindent \textbf{(iii) Co-citation network ($CCN$)}: In this directed network, each author is considered as a node and two authors are connected by an edge ($a_{i} \longrightarrow a_{j}$) if author $a_i$ has cited an article authored by $a_j$. There is bidirectional edge ($a_{i} \longleftrightarrow a_{j}$) if author $a_i$ and author $a_j$ cites each other.

For our experiments, we consider the authors who have submitted their paper to the Journal of High Energy Physics (JHEP) between 1997 to 2015. We consider approx. 29k papers and more than 24k authors. We also have approx. 70k unique review reports.
\subsection{Key results}
A nuanced analysis shows that authors in the class $ACC_{high}$ usually receive more citations than the other two categories. We also note that papers of the $ACC_{low}$ authors receive more citation if they coauthored with $ACC_{high}$ authors in some paper. $ACC_{high}$ authors always receive more positive reviews than the other two categories. An intriguing observation is that the set of referees and editors to whom the papers of the $ACC_{high}$ class are assigned are found to be less diverse than the other two classes. The $ACC_{high}$ authors are more `central' in all the networks. We finally make early predictions of the $ACC$ category of an author and obtain 0.82 - 0.95 precision and 0.82 - 0.91 recall. In a followup discussion we narrate how apart from the author characteristics, the peer review system itself can potentially facilitate discrimination in the editing and the reviewing process of papers in the three categories which could reinforce the distinction between the authors of these categories and calls for further investigation by the system admins.

\subsection{Outline}
The rest of the paper is organised as follows.  Section~\ref{sec:datasets} describes the dataset used in this paper. Section~\ref{sec:categorization} details the method for author categorization. Section~\ref{sec:authors_feature} and~\ref{sec:peer_review_features} demonstrate the author profile features and peer review based features respectively. In section~\ref{sec:graph_based}, we discuss the network features of the three category of authors. In section~\ref{prediction} we predict the category of the authors. In section~\ref{sec:catwise_authors_editor_reviewer_analysis} discuss the potential role of the peer review system in enhancing the distinction among the three categories of authors. Section~\ref{sec:related_work} presents a brief literature review. Finally, we conclude in section~\ref{conc}.

\section{Dataset Description}\label{sec:datasets}
In our article, we consider papers submitted to the Journal of High Energy Physics (JHEP)\footnote{https://jhep.sissa.it/jhep/} in between 1997 and 2015. JHEP is one of the leading journals in the domain of high energy physics. In JHEP, the identity of the referee remains confidential. This dataset contains a total of 28871 papers, where the number of accepted and rejected papers are 20384 and 6190 respectively. We also have 70000 unique peer review reports. For each paper we have the title, author names, broad topics that the paper is on, publication date (in case it was accepted) and the number of citations for the accepted papers. In addition, this dataset contains the review text, number of review rounds, editor and reviewer ids (anonymised) of each paper. We also have the citation link among the papers. For the rejected papers, we collected the arXiv\footnote{http://arxiv.org} id using the Inspire\footnote{https://inspirehep.net} search engine. We consider the cumulative number of citations obtained at the end of 2015. We present a brief statistics of the dataset in Table~\ref{tab:dataset}.
\vspace{-0.2cm}
\begin{table}[tbhp]
\centering
\caption{\label{tab:dataset}Dataset description.}
\begin{adjustbox}{width=0.35\textwidth}
\begin{tabular}{|c|c|c|c|} \hline
{\bf Basic Information} & {\bf Count} \\ \hline
\#papers & 26574\\ \hline
\#unique authors & 24868 \\ \hline
\#papers (accepted) & 20384 \\ \hline
\#papers (rejected) & 6190 \\ \hline
Average \#citations (accepted papers) & 31.88 \\ \hline
Average \#citations (rejected papers)  & 9.45 \\ \hline
\end{tabular}
\end{adjustbox}
\end{table}

\section{Author Categorization}
\label{sec:categorization}
\begin{figure}[!tbh]
\centering
\setlength\tabcolsep{0pt}
\renewcommand{\arraystretch}{0.8}%
\begin{tabular}{@{}c@{}c@{}c@{}c@{}}
  \includegraphics[width=.45\textwidth]{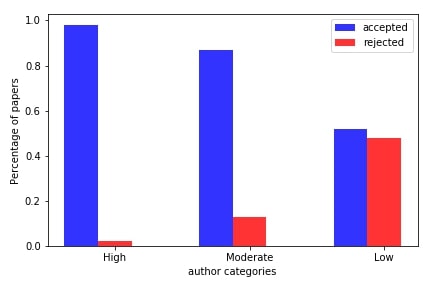}\\
\end{tabular}
\caption{Percentage of accepted and rejected papers of $ACC_{high}$ (High), $ACC_{mid}$ (Moderate) and $ACC_{low}$ (Low) authors. }  
\label{fig:categorywise_percentageOfPapers}
\end{figure}

In this section, we categorize authors' profile into three categories based on their articles' acceptance rate ($ACC$) -- (i) authors with high acceptance ($ACC_{high}$) (ii) authors with moderate acceptance ($ACC_{mid}$) (iii) authors with low acceptance ($ACC_{low}$). Acceptance rate of an author is calculated as the ratio of the number of papers accepted to number of papers submitted by that author. We calculate article acceptance rate of each author for every year.  In case of $ACC_{high}$ category, we consider only those authors who have high acceptance rate $(>0.7)$ in at least 70\% of the years over all the years. $ACC_{low}$ category contains authors who have very low acceptance rate $(<0.4)$ in at least 80\% of the years. We keep the rest of the authors (not falling in the other two categories) in $ACC_{mid}$ category. Statistics of the unique authors are given in Table~\ref{tab:category_stat}. 
The number of accepted and rejected papers in each class are noted in Figure~\ref{fig:categorywise_percentageOfPapers}. The papers of authors in the $ACC_{high}$ class almost always get accepted.

\begin{table}[tbhp]
\centering
\caption{Statistics of author categorization.}\label{tab:category_stat}
\begin{adjustbox}{width=0.3\textwidth}
\begin{tabular}{|c|c|} \hline
{\bf Author Categories} & {\bf \#Authors} \\ \hline
$ACC_{high}$ & 3688 \\ \hline
$ACC_{mid}$ & 10359 \\ \hline
$ACC_{low}$ & 9644 \\ \hline
\end{tabular}
\end{adjustbox}
\end{table}

\begin{figure*}[!tbh]
\setlength\tabcolsep{0pt}
\renewcommand{\arraystretch}{0.8}%
\begin{tabular}{@{}c@{}c@{}c@{}c@{}}
  \includegraphics[width=0.45\hsize, height=7.5cm]{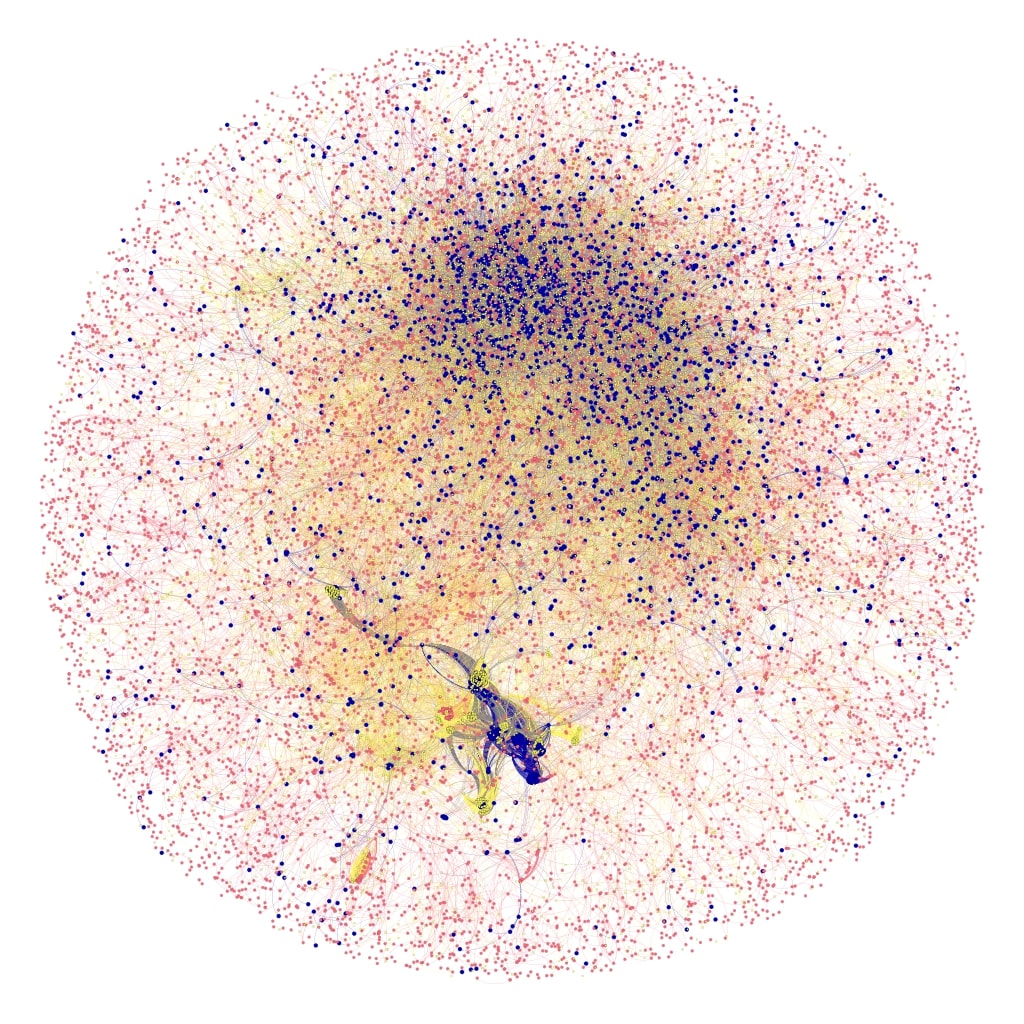}
  \includegraphics[width=0.45\hsize, height=7.5cm]{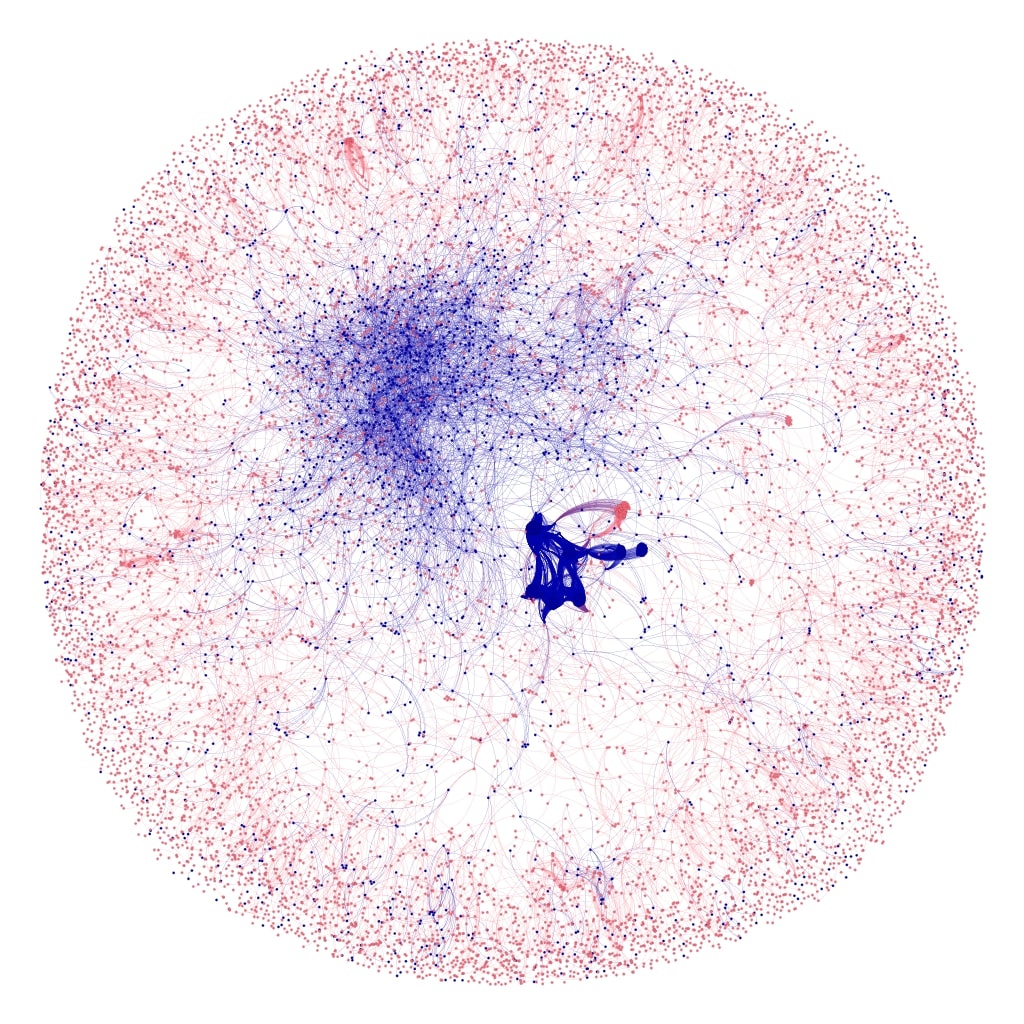} \\ 
  \includegraphics[width=5cm,height=0.5cm,right]{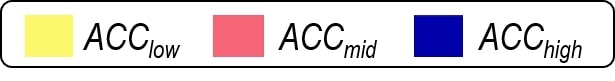}
\end{tabular}
\caption{(Left) This collaboration network includes $ACC_{high}$, $ACC_{mid}$ and $ACC_{low}$ authors. (Right) This collaboration network includes $ACC_{high}$ and $ACC_{mid}$ authors only for better visualisation.}  
\label{fig:coauthorship}
\end{figure*}

\section{Author profile based features ($AP_f$)}
\label{sec:authors_feature}

\subsection{Citation index ($C_{ind}$)}
Citation count of each author is computed by considering the total number of citations an author received in their active period. For each category, we define citation index as the \textit{standard deviation of citation counts} of all the authors. We compute $C_{ind}$ for three categories $ACC_{high}$, $ACC_{mid}$ and $ACC_{low}$ (see Figure~see Figure~\ref{fig:author_features}(d)). There is a stark difference in the values of $C_{ind}$ among the three categories. Authors in the class $ACC_{high}$ have low $C_{ind}$ (approx. 50) whereas authors in the class $ACC_{low}$ have high $C_{ind}$ (approx. 101). Thus, the citation counts in the class $ACC_{high}$ are far more uniform across the authors compared to the $ACC_{low}$ class. 

\subsection{Experience ($E_{cnt}$)}
Experience of an author is defined in terms of number of papers he has published. For each category, we compute experience of all the authors and consider the mean of these $E_{cnt}$s. We observe $ACC_{high}$ has highest mean experience (see Figure~\ref{fig:author_features}(a)). $ACC_{mid}$ has moderate value of mean experience whereas $ACC_{low}$ has very low mean experience (see Figure~\ref{fig:author_features}(a)). From this, it is clear that $ACC_{low}$ category contains those authors who are either new in research or has very few publications. 
\if{0}
\begin{table}[tbhp]
\centering
\caption{Mean experience ($E_{cnt}$) of authors in different categories}\label{tab:experience}
\begin{adjustbox}{width=0.3\textwidth}
\begin{tabular}{|c|c|} \hline
{\bf Author Categories} & {\bf Mean Experience} \\ \hline
$ACC_{high}$ & 8.49 \\ \hline
$ACC_{mid}$ & 5.35 \\ \hline
$ACC_{low}$ & 1.72 \\ \hline
\end{tabular}
\end{adjustbox}
\end{table}
\fi
\subsection{Topic diversity ($T_{div}$)}
We consider a topic set for each author. This topic set contains all the topics on which an author published their papers. For each author we compute \textit{topic ratio} as the ratio of the total number of topics on which he/she has written a paper to the total number of papers he/she published. For each category, we consider mean over \textit{topic ratio} of all the authors to compute topic diversity ($T_{div}$) (see Figure~\ref{fig:author_features}(e)). Interestingly, $ACC_{high}$ category authors have less $T_{div}$ (1.03) than the other two categories ($ACC_{mid}$ has 1.36 and $ACC_{low}$ has 1.57). We observe that $ACC_{low}$ category authors publish papers on a lot of topics whereas $ACC_{high}$ authors focus on a relatively less number of topics and publish a large number of papers in those topics.

\subsection{$h$-index ($H_{ind}$)}
The $h$-index~\cite{Hirsch:2005} is defined as the maximum value of $h$ such that an author has published $h$ papers that have each been cited at least $h$ times. For all the three categories, we consider mean of the $H_{ind}$ of all the authors. From Figure~\ref{fig:author_features}(b), it is clear that $ACC_{high}$ have very high mean $H_{ind}$ compared to the other two categories. Thus the $ACC_{high}$ class usually comprises the high impact authors.

\subsection{Team size ($TS$)}
\begin{figure*}[!tbh]
\centering
\setlength\tabcolsep{0pt}
\renewcommand{\arraystretch}{0.8}%
\begin{tabular}{@{}c@{}c@{}c@{}c@{}}
  \includegraphics[width=1\hsize]{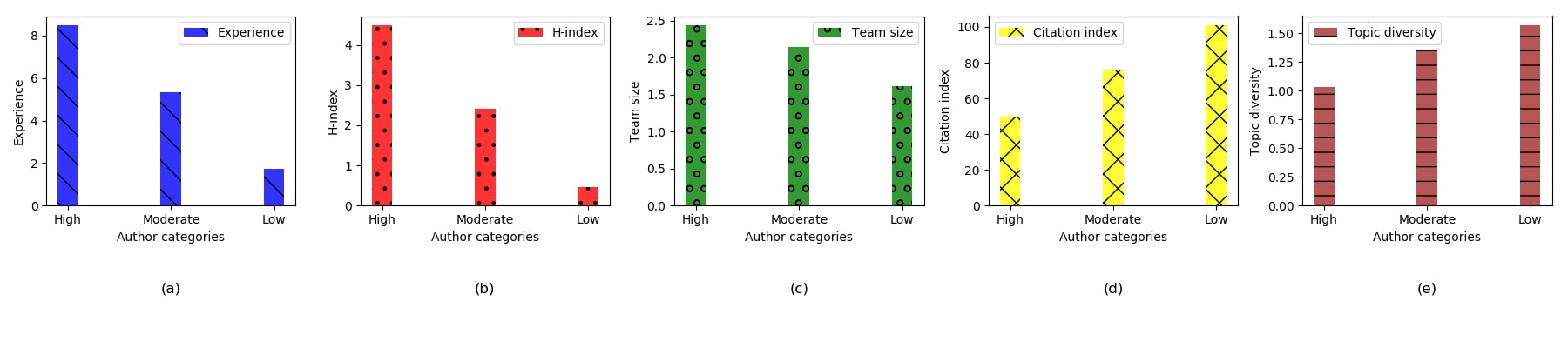} \\
\end{tabular}
\caption{(a) The mean experience ($E_{cnt}$) of $ACC_{high}$ (High), $ACC_{mid}$ (Moderate) and $ACC_{low}$ (Low). (b) The mean $h$-index for the three categories. (c) The mean team size ($TS$) for the three categories. (d) Citation index ($C_{ind}$) of the three categories. (e) Topic diversity ($T_{div}$) for the three categories.}  
\label{fig:author_features}
\end{figure*}

Team size of an author is calculated as the number of contributing co-authors averaged across all the papers that the particular author has written. We examine mean team size for each category (see Figure~\ref{fig:author_features}(c)). $ACC_{high}$ and $ACC_{mid}$ authors have mean team sizes of 2.44 and 2.15. The typical team sizes for both these classes are very similar. On the other hand, we find that the mean team size of $ACC_{low}$ is $\sim1.61$ which is quite low compared to the other two classes.

\section{Peer review text based features ($PF_f$)}
\label{sec:peer_review_features}
\subsection{Sentiment of review text ($SNT_{r}$)}
We compute the sentiment score $[-1,1]$ of each review text for each paper\footnote{https://textblob.readthedocs.io/en/dev/}. For every author we compute the average review sentiment across all the papers (s)he has written. For every class, we take the mean of these average values across all the authors in that class (see Figure~\ref{fig:review_features}(a)). Among the three classes, the review text bears the highest positive sentiment (0.15) in the $ACC_{high}$ class. This is followed by $ACC_{mid}$ class where the overall sentiment is 0.05. Finally, the review texts corresponding to the $ACC_{low}$ class indicate the presence of high negative sentiment $(-0.26)$. 

\if{0}
\begin{figure*}[!tbh]
\centering
\setlength\tabcolsep{0pt}
\renewcommand{\arraystretch}{0.8}%
\begin{tabular}{@{}c@{}c@{}c@{}}
  \includegraphics[width=.33\hsize]{Figures/BiHighAccepted.jpg} &
  \includegraphics[width=.33\hsize]{Figures/BiMediumAccepted.jpg}& 
  \includegraphics[width=.33\hsize]{Figures/BiLowAccepted.jpg}\\
  (a) $ACC_{high}$ accepted papers & (b) $ACC_{mid}$ accepted papers & (c) $ACC_{low}$ accepted papers\\
  \includegraphics[width=.33\hsize]{Figures/BiHighRejected.jpg} &
  \includegraphics[width=.33\hsize]{Figures/BiMediumRejected.jpg}& 
  \includegraphics[width=.33\hsize]{Figures/BiLowRejected.jpg}\\
  (d) $ACC_{high}$ rejected papers & (e) $ACC_{mid}$ rejected papers & (f) $ACC_{low}$ rejected papers\\
\end{tabular}
\caption{Wordclouds of frequent bi-grams in review text for accepted and rejected papers. (a) Review text of accepted papers of $ACC_{high}$ is positive and review contains positive bi-grams like \em{well written}, \em{I recommend} etc. (b) Review text of accepted papers of $ACC_{mid}$ contains positive words like \em{I recommend}, \em{recommended paper} but it also contain few negative bi-grams such as \em{can not}, \em{not clear} etc.}
\label{fig:review_text_emotion}
\end{figure*}
\fi
\subsection{Length of review text ($L_{r}$)}
Length of review text is computed as the number of words present in the review text except stop-words (see Figure~\ref{fig:review_features}(b)). Surprisingly, we find that $ACC_{high}$ category receive relatively lengthier reviews (2368) compared to $ACC_{low}$ (1305). It is therefore quite clear that papers in the $ACC_{high}$ class typically receive more detailed feedback from the referees compared to the $ACC_{low}$ class. 

\subsection{Reviewer diversity ($R_{div}$)}\label{ref_div}
We use Shannon index~\cite{Spellerberg:2003} to calculate the reviewer diversity. For each author in a particular category, we extract the reviewer ids of all his/her published papers and add it to a global list. Thus we have three global lists for each of the three categories. Next, for each category, we compute the entropy of this global list. Let the size of the global list for a category be $N$ and let the number of occurrences of a reviewer $r_i$ in the list be $f_i$. Then the entropy would be $-\sum_{\forall{i}}\frac{f_i}{N}log(\frac{f_i}{N})$. If the value of this entropy is low then this would mean that the number of reviewers to whom the papers of a class go for review are very limited. In contrast, if this value is high for a class then it would mean that many reviewers are assigned as referees for the papers in the class (see Figure~\ref{fig:review_features}(c)). Surprisingly, we notice that $ACC_{high}$ has less reviewer diversity $(\sim 6.83)$ than the other two categories. $ACC_{mid}$ and $ACC_{low}$ categories have reviewer diversity 7.36 and 7.34 respectively. This possibly indicates that for the $ACC_{high}$ class the set of referees are relatively more fixed and papers of authors from this group usually go to other peer authors (in the role of referees) mostly from this group itself for a review. This, we believe, is a sign of unhealthy reviewing practice. We shall discuss more about this in section~\ref{sec:catwise_authors_editor_reviewer_analysis}.

\subsection{Editor diversity ($Ed_{div}$)}\label{ed_div}
\begin{figure*}[!tbh]
\centering
\setlength\tabcolsep{0pt}
\renewcommand{\arraystretch}{0.8}%
\begin{tabular}{@{}c@{}c@{}c@{}c@{}}
  \includegraphics[width=1\hsize]{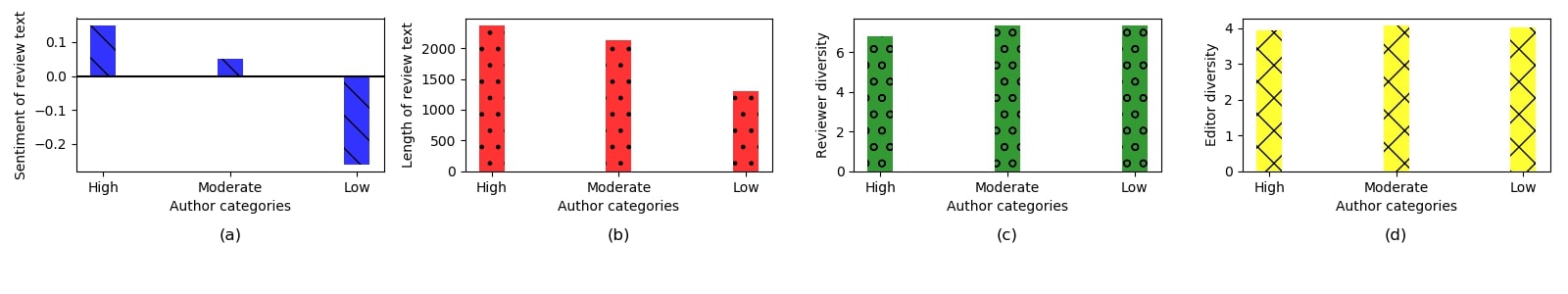} \\
\end{tabular}
\caption{(a) The sentiment of review text ($SNT_{r}$) of $ACC_{high}$ (High), $ACC_{mid}$ (Moderate) and $ACC_{low}$ (Low). (b) The length of the review text ($L_{r}$) for the three categories. (c) Reviewer diversity ($R_{div}$) of the three categories. (d) Editor diversity ($Ed_{div}$) for the three categories.}  
\label{fig:review_features}
\end{figure*}

Once again we use Shannon index~\cite{Spellerberg:2003} to calculate editor diversity. We compute this metric exactly as $R_{div}$ with the exception that here the three global lists are composed of editor ids to whom the papers are assigned (as opposed to reviewer ids in the previous case). Here also we observe that editor diversity of $ACC_{high}$ is quite low $\sim 3.94$; on the other hand, the editor diversity of $ACC_{mid}$ and $ACC_{low}$ classes are relatively higher $\sim 4.07$ and $\sim 4.01$ respectively (see Figure~\ref{fig:review_features}(d)). It seems that the same set of editors handle the papers of the $ACC_{high}$ class.

\subsection{Linguistic quality indicator ($LQI$)}
Here we analyze the different emotions (positive, optimism, cheerfulness, confusion and contentment) reflected by each word present in the review text\footnote{https://github.com/Ejhfast/empath-client}. Then we take the mean of the emotion values of words present in a particular review text and average it over all authors in a class. We find quite a few interesting results. There are more positive emotion words in the review texts of the $ACC_{high}$ class (0.018) compared to the $ACC_{low}$ class (0.015). Further, there are more optimism related words in the review texts of the $ACC_{high}$ class (0.01) compared to the $ACC_{low}$ class (0.004). There are more cheerfulness related words present in the review texts of the $ACC_{high}$ class (0.0017) compared to the $ACC_{low}$ class (0.0014). There are less confusion words in the review texts of the $ACC_{high}$ class (0.0026) compared to the $ACC_{low}$ (0.0036) class. Last, there are more contentment related words in the review texts of the $ACC_{high}$ class (0.0079) compared to the $ACC_{low}$ class (0.0059).

\section{Network analysis based features ($NE_{f}$)}
\label{sec:graph_based}

\begin{figure}[!tbh]
\centering
\setlength\tabcolsep{0pt}
\renewcommand{\arraystretch}{0.8}%
\begin{tabular}{@{}c@{}c@{}c@{}c@{}}
  \includegraphics[width=0.8\hsize,height=8cm]{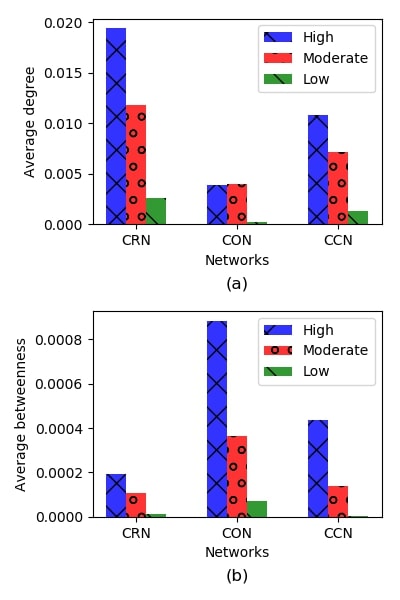} \\
\end{tabular}
\caption{(a) The average degree centrality of $ACC_{high}$ (High), $ACC_{mid}$ (Moderate) and $ACC_{low}$ (Low) for the three networks ($CRN$, $CON$ and $CCN$). (b) The average betweenness centrality of $ACC_{high}$ (High), $ACC_{mid}$ (Moderate) and $ACC_{low}$ (Low) for three networks.}  
\label{fig:deg_bet_cc_plot}
\end{figure}

\if{0}
\subsection{\textbf{Density}}
Density~\cite{} of a network can be defined as the degree of connections among nodes. Density is calculated as the number of edges present in the network divided by number of possible edges. To understand how well each category authors are connected among themselves, we calculate density of the whole network as well as for each category induced network.

\subsection{\textbf{Assortativity}}
Assortativity~\cite{Newman:2003} is defined as a metric to measure the preference for nodes to attach to others that are similar in some way. For example, a network is called assortative if high degree nodes are connected to high degree nodes and vice versa. We compute assortativity of overall and induced co-reviewer graph.  The overall assortativity of whole co-reviewer graph is 0.63.

\subsection{\textbf{Reciprocity}}
Reciprocity is defined as a ratio of number of connections that are mutually linked to total number of possible connections. This measure is for directed network. We computed reciprocity for co-citation network and its induced subgraphs.

\subsection{Degree centrality}
Degree centrality of a node is defined as the ratio of the number of neighbors it has and the total number of possible neighbors. The node with high degree centrality is most important node in the network. We compute degree centrality for nodes of three different networks.{\color{blue}Add the plots}

\subsection{Betweenness centrality}
Betweenness centrality~\cite{Freeman:1977} of a node is defined by ratio of the number of shortest path passes through the node to the total number of shortest path from source to destination. The node with high betweenness centrality has more information than the other nodes. We compute betweenness centrality for all the three types of networks.

\subsection{Clustering coefficient}
Clustering coefficient~\cite{Giorgio:2007} of a node is computed as the ratio of the number of edges among its neighbors to the number of possible edges among its neighbors. Clustering coefficient shows how well a node's neighbors are connected among themselves. We also compute clustering coefficient of nodes for each network.

\subsection{Closeness centrality}
Closeness centrality~\cite{Freeman:1978} of a node is the inverse of the sum of all the shortest path between the node and all other nodes in the network. The node with highest closeness centrality is most central node in the network and closest to all other nodes.

\subsection{Page rank}
Page rank is an algorithm which estimates importance of a node in the network. The main assumption of this algorithm is the more important nodes likely to have more incoming edges than others. We computed page rank on $CCN$ and $CRN$ network.
\fi

In this section, we study the properties of the three different networks in details.

\subsection{Analysis of the co-reviewer network ($CRN$)}

Recall that in a co-reviewer network each node corresponds to an author and two authors are connected if their papers have been co-reviewed by the same referee. We run series of analysis on this network to investigate the differences between the three categories.

\subsubsection{Centrality measures}
Here we compute four centrality measures of the whole co-reviewer network.

\noindent{\bf Degree centrality}: We compute the average degree centrality of the authors (see Figure~\ref{fig:deg_bet_cc_plot}(a)) for each category. We observe that the average degree centrality of authors of $ACC_{high}$ category is high (0.019) whereas the average degree centrality of the authors for $ACC_{mid}$ ($\sim 0.011$) and $ACC_{low}$ ($\sim 0.002$) are relatively lower. (see Figure~\ref{fig:deg_bet_cc_plot}(a)). 

\noindent{\bf Betweenness centrality}: We compute the average betweenness centrality of the authors of each category. The average betweenness centrality (see Figure~\ref{fig:deg_bet_cc_plot}(b)) of $ACC_high$ category is marginally higher ($\sim 0.00019$) than the other two categories. 


\noindent{\bf Closeness centrality}: We calculate the average closeness centrality of authors for each category. The average closeness centrality (see Figure~\ref{fig:clos_pg_plot}(a)) of $ACC_high$ category is higher ($\sim 0.362$) than the other two categories. 

\noindent{\bf PageRank}: We calculate the average PageRank score of the authors for each category. The average PageRank (see Figure~\ref{fig:clos_pg_plot}(b)) of $ACC_high$ category is marginally higher ($\sim 0.0000719$) than the other two categories. 


\subsubsection{Core periphery analysis}
Here we perform a $k$-shell decomposition of the network and inspect four different shells -- the innermost ($k=180$), the inner-mid ($k=140$), the outer-mid ($k=90$) and the outermost ($k=1$). As noted in Table~\ref{tab:core_periphery}, we observe that the innermost and inner-mid shells contain a larger fraction of nodes from the $ACC_{high}$ and $ACC_{mid}$ classes compared to the $ACC_{low}$ class. In contrast, the outermost shell contains the largest fraction of nodes from the $ACC_{low}$ class. 

\begin{table}[tbhp]
\centering
\caption{Core periphery analysis of the co-reviewer network.}\label{tab:core_periphery}
\begin{adjustbox}{width=0.45\textwidth}
\begin{tabular}{|c|c|c|c|c|} \hline
{\bf Shell}&{\bf \# Authors} & {\bf \% $ACC_{high}$} & {\bf \% $ACC_{mid}$} & {\bf \% $ACC_{low}$}\\ \hline
Innermost (180) & 167 & \cellcolor{green!20}29.9 & \cellcolor{green!20}49.1 & 20.3 \\ \hline
Inner-mid (140)& 37 & \cellcolor{green!20}13.5 & \cellcolor{green!20}78.3 & 8.1 \\ \hline
Outer-mid (90) & 116 & 11.2 & 48.2 & 36.2 \\ \hline 
Outermost (1) & 227 & 7 & 14.5 & \cellcolor{red!20}62 \\ \hline
\end{tabular}
\end{adjustbox}
\end{table}
\subsubsection{Induced co-reviewer network}
Here we construct three induced co-reviewer networks comprising the authors in the classes $ACC_{high}$, $AC_{mid}$ and $AC_{low}$ respectively. 

\noindent{\bf Density}: We calculate the density of each induced graph to observe how densely the authors are connected among themselves through the common reviewers. Density of the $ACC_{high}$ induced graph is higher (0.047) than others. Density of $ACC_{mid}$ and $ACC_{low}$ are 0.016 and 0.001 respectively. 

\noindent{\bf Assortativity coefficient}: We compute the assortativity coefficient of the three induced networks. While this coefficient for the $ACC_{high}$ induced graph is as high as 0.82, the same for the $ACC_{mid}$ and the $ACC_{low}$ induced graphs are 0.66 and 0.24 respectively. This indicates that the $ACC_{high}$ induced graph is much more homophilic compared to the other two graphs.

\noindent{\bf Edge transitions}: We finally study the edge transitions among the three induced graphs, i.e., given a pair of induced graphs we find the fraction of edges going from one of them to the other from the original co-reviewer network. We find that $ACC_{high}$ and $ACC_{mid}$ share almost 34.7\% edges whereas $ACC_{high}$ and $ACC_{low}$ share only 4.3\% edges. The fraction of edges between $ACC_{mid}$ and $ACC_{low}$ is around 10.4\%. 
\vspace{-0.3cm}
\begin{figure}[!tbh]
\centering
\setlength\tabcolsep{0pt}
\renewcommand{\arraystretch}{0.8}%
\begin{tabular}{@{}c@{}c@{}c@{}c@{}}
  \includegraphics[width=0.8\hsize,height=8cm]{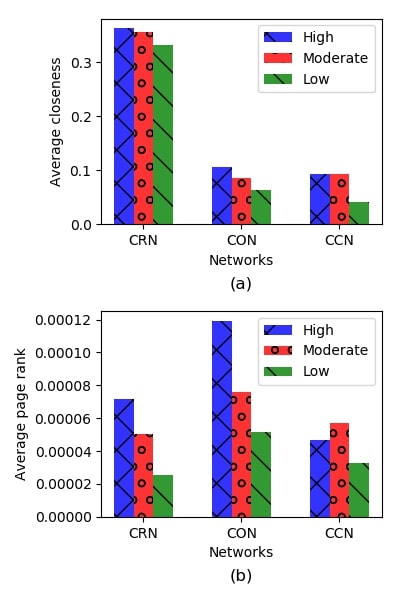} \\
\end{tabular}
\caption{(a) The average closeness centrality of $ACC_{high}$ (High), $ACC_{mid}$ (Moderate) and $ACC_{low}$ (Low) for three networks ($CRN$, $CON$ and $CCN$). (b) The average PageRank of $ACC_{high}$ (High), $ACC_{mid}$ (Moderate) and $ACC_{low}$ (Low) for three networks.}  
\label{fig:clos_pg_plot}
\end{figure}
\vspace{-0.2cm}

\subsection{Analysis of co-citation network ($CCN$)}

Recall the the co-citation network has authors as its nodes and there is an edge from author $a_i$ to $a_j$ if $a_i$ cites a paper of $a_j$. If both $a_i$ and $a_j$ cite each other in some of their papers then there is a bidirectional edge between them.

\subsubsection{Centrality measures}
We compute four centrality measures in the co-citation network.

\noindent{\bf Degree centrality}: We compute the average degree centrality of the authors (see Figure~\ref{fig:deg_bet_cc_plot}(a)) for each category. We observe that the average degree centrality of the authors of $ACC_{high}$ category is high compared to the average degree centrality of authors for $ACC_{mid}$ and $ACC_{low}$ categories. 

\noindent{\bf Betweenness centrality}: We compute the average betweenness centrality of the authors for each category. The average betweenness centrality (see Figure~\ref{fig:deg_bet_cc_plot}(b)) of the authors of $ACC_{high}$ category is marginally higher ($\sim 0.0004$) than the other two categories. 


\noindent{\bf Closeness centrality}: We calculate the average closeness centrality of the authors of each category. The average closeness centrality (see Figure~\ref{fig:clos_pg_plot}(a)) of $ACC_{high}$ category authors is higher ($\sim 0.092458$) than the other two categories. 

\noindent{\bf PageRank}: We calculate the average PageRank score of the authors of each category. The average PageRank (see Figure~\ref{fig:clos_pg_plot}(b)) of $ACC_{high}$ and $ACC_{mid}$ categories are marginally higher than the $ACC_{low}$ category. 

\subsubsection{Induced co-citation network}
 Here again we construct three induced co-citation networks comprising the authors from the three classes -- $ACC_{high}$, $ACC_{mid}$ and $ACC_{low}$. 
 
 \noindent{\bf Cross citations}: We find the fraction of citations running in between the classes. Notably, the largest fraction of citation edges run between $ACC_{high}$ and $ACC_{mid}$ induced graphs (45\%). Fraction of citation edges running between $ACC_{high}$ and $ACC_{low}$ induced graphs, on the other hand, is the least (1\%).
 
 \noindent{\bf Self citations}: Fraction of citation edges running within the $ACC_{high}$ induced graph is the highest $(\sim 33.2\%)$. This fraction for the $ACC_{mid}$ and $ACC_{low}$ are 17.9\% and 0.4\% respectively.
 
\noindent{\bf Reciprocity}: We compute the reciprocity within and across all the three induced networks. Reciprocity within the $ACC_{high}$ induced network is the highest (0.61); reciprocity in the $ACC_{mid}$ induced network is 0.20 and the same for the $ACC_{low}$ induced network is 0.08 which is the least among the three.

Reciprocity in between $ACC_{high}$ and $ACC_{mid}$ induced networks is 0.34 which is higher than between $ACC_{mid}$ and $ACC_{low}$ (0.11) as well as $ACC_{high}$ and $ACC_{low}$ (0.12).


\subsubsection{$ACC_{low}$ authors that are cited by $ACC_{high}$ authors}
Although a rare case, here, we observe how the citations coming from the $ACC_{high}$ authors affect the fate of the papers written by the $ACC_{low}$ authors. We separately consider those papers which are cited by $ACC_{high}$ authors and observe the author characteristics of such papers. We find that the mean citation of papers written by $ACC_{low}$ authors and cited by $ACC_{high}$ authors is roughly double $(\sim 57.91)$ the mean citation of papers $(\sim 28.9)$ written by $ACC_{low}$ authors that are never cited by the $ACC_{high}$ authors. 

We further notice that the mean citation of those $ACC_{low}$ authors $(\sim 50.07\%)$ whose papers are cited by $ACC_{high}$ authors is higher than the mean citation of the other $ACC_{low}$ authors $(\sim 28.94\%)$.

\if{0}
\begin{figure}[!tbh]
\centering
\setlength\tabcolsep{0pt}
\renewcommand{\arraystretch}{0.8}%
\begin{tabular}{@{}c@{}c@{}c@{}}
  \includegraphics[width=1\hsize]{Figures/cdf_plot.png} \\
\end{tabular}
\caption{Cumulative distribution function (CDF) of citations received by the authors of $ACC_{high}$,$ACC_{mid}$ and $ACC_{low}$}
\label{fig:log_citation_count}
\end{figure}
\fi

\subsubsection{$ACC_{low}$ authors cited by $ACC_{mid}$ authors}
In this section, we investigate the characteristics of those $ACC_{low}$ authors whose papers are cited by $ACC_{mid}$ authors. Once again, we observe that the mean citation of papers $(\sim 50.46)$ written by $ACC_{low}$ authors and cited by $ACC_{mid}$ authors is much higher than the mean citation of papers $(\sim 28.9)$ written by $ACC_{low}$ authors but never cited by the $ACC_{mid}$ authors.

\subsection{Analysis of collaboration network ($CON$)}

Recall that the in the collaboration network each node is an author and two authors are connected if they have co-authored a paper together. We present a visualisation of the collaboration network in Figure~\ref{fig:coauthorship}. The left sub-figure shows the authors in the three categories as nodes of different colours. The blue nodes correspond to the authors in the $ACC_{high}$ category, the red nodes correspond to the authors in the $ACC_{mid}$ category and the yellow nodes correspond to the authors in the $ACC_{low}$ category. The blue nodes are concentrated mostly in the center of the network while the red and the yellow nodes are scattered all across the network. This is more clear when we draw the network of the authors corresponding to the $ACC_{high}$ and the $ACC_{mid}$ category. The blue nodes are largely concentrated at the center of the network.

\subsubsection{Centrality measures}
We compute four centrality measures from the collaboration network.

\noindent{\bf Degree centrality}: We compute the average degree centrality of the authors (see Figure~\ref{fig:deg_bet_cc_plot}(a)) for each category. We observe that the average degree centrality of the authors in the $ACC_{high}$ category is higher (0.0039) than the average degree centrality of the authors in the other two categories. 

\noindent{\bf Betweenness centrality}: We compute the average betweenness centrality of the authors of each category. The average betweenness centrality (see Figure~\ref{fig:deg_bet_cc_plot}(b)) of $ACC_{high}$ category is higher ($\sim 0.00088$) than the other two categories. 


\noindent{\bf Closeness centrality}: We calculate the average closeness centrality of the authors of each category. The average closeness centrality (see Figure~\ref{fig:clos_pg_plot}(a)) of $ACC_high$ category is higher ($\sim 0.105$) than the other two categories. 

\noindent{\bf PageRank}: We calculate the average PageRank score of the authors of each category. The average PageRank (see Figure~\ref{fig:clos_pg_plot}(b)) of $ACC_high$ category is marginally higher ($\sim 0.000119$) than the other two categories. 

\subsubsection{Class wise collaborations}
The fraction of collaboration edges between the $ACC_{high}$ and $ACC_{mid}$ authors is 38.9\% which is much higher than either the fraction of collaboration edges between $ACC_{mid}$ and $ACC_{low}$ authors (1.0\%) or $ACC_{high}$ and $ACC_{low}$ authors (0.2\%). 

On the other hand, the fraction of collaboration edges within the $ACC_{high}$ authors is 26.4\%, while this is 31.3\% for the $ACC_{mid}$ authors and 0.7\% for the $ACC_{low}$ authors.

\subsubsection{$ACC_{low}$ authors collaborating in papers primarily written by $ACC_{high}$ authors}

In this section, we focus on those $ACC_{low}$ authors who get a chance to collaborate with $ACC_{high}$ authors. In particular, we consider those papers which are written by a mix of 20\% $ACC_{low}$ authors and 80\% $ACC_{high}$ authors (i.e., papers predominantly written by authors with high acceptance ratio).

We compute various features discussed earlier for this 20\% $ACC_{low}$ authors when they write papers with $ACC_{high}$ authors and when they write papers without them. The feature values are noted in Table~\ref{tab:collaborate_ACC_high_80}. Collaborations with the $ACC_{high}$ authors seems to heavily benefit the $ACC_{low}$ authors in terms of accrued citations as well as review sentiments obtained from the referees.


\begin{table}[tbhp]
\centering
\caption{Properties of $ACC_{low}$ authors who collaborate with a high number $ACC_{high}$ authors.}\label{tab:collaborate_ACC_high_80}
\begin{adjustbox}{width=0.48\textwidth}
\begin{tabular}{|c|c|c|} \hline
{\bf Features}&{\bf Collaborated} & {\bf Not collaborated}\\
& {\bf  with $ACC_{high}$} & {\bf  with $ACC_{high}$ } \\ \hline
Mean \#papers & 1.1 & 1.9 \\ \hline
Team size ($TS$) &4.3 & 3.1 \\ \hline
Citation ($C_{cnt}$) & \cellcolor{green!20}30 & 12 \\ \hline 
Review text sentiment ($SNT_r$) & \cellcolor{green!20}0.23 & -0.13 \\ \hline
\end{tabular}
\end{adjustbox}
\end{table}

\subsubsection{$ACC_{high}$ authors collaborating in papers primarily written by $ACC_{low}$ authors}
In this section, we analyze such cases where papers are written by 80\% $ACC_{low}$ and 20\% $ACC_{high}$ authors. We analyze profile features of these 80\% $ACC_{low}$ authors when they write papers with $ACC_{high}$ authors as well as when they write without them. Table~\ref{tab:collaborate_ACC_high_20} enumerates the important features and shows that even having a small fraction of $ACC_{high}$ authors in their paper can increase the citation count and reduce the negative sentiment in the reviews of the $ACC_{low}$ authors.

\begin{table}[tbhp]
\centering
\caption{Analysis of $ACC_{low}$ authors who collaborate with a low number of $ACC_{high}$ authors.}\label{tab:collaborate_ACC_high_20}
\begin{adjustbox}{width=0.48\textwidth}
\begin{tabular}{|c|c|c|} \hline
{\bf Features}&{\bf Collaborated} & {\bf Not collaborated}\\
& {\bf  with $ACC_{high}$} & {\bf  with $ACC_{high}$ } \\ \hline
Mean \#papers & 1.0 & 1.8 \\ \hline
Team size ($TS$) &4.12 & 2.85 \\ \hline
Citation ($C_{cnt}$) & \cellcolor{green!20}53.7 & 27.7 \\ \hline 
Review text sentiment ($SNT_r$) & \cellcolor{green!20}-0.39 & -0.54 \\ \hline
\end{tabular}
\end{adjustbox}
\end{table}

\section{Author category prediction}\label{prediction}
\subsection{Classification model}
In our classification model, we consider $AP_{f}$, $PF_{f}$ and $NE_{f}$ features for the first three years of career of each author as the training data. For example, if an author published his first paper in 1996 then we consider papers published in between 1996 and 1998 for training purpose. We compute all the features of an author based on the first three years of career information. For testing, we leave a gap of two years to prevent any data leakage. After five years, we predict their category. We use two different classifiers -- XGBoost~\cite{Chen:2016} and random forest~\cite{Breiman:2001}. In order to evaluate the model, we compute class wise precision and recall. In addition, we also compute F1-score. We calculate precision as the fraction of authors who are correctly classified out of all the predicted authors. Recall is the fraction of relevant authors correctly classified by the classifier.

\noindent{\bf Features}: We use the author profile features ($AP_{f}$), peer review based features ($PF_{f}$) as well as network features ($NE_{f}$). 

\noindent{\bf Results}: The class wise precision and recall for the XGBoost model are noted in Table~\ref{tab:XGBoost_precision_recall}. The F1-score for the model is 0.84. The confusion matrix is tabulated in Table~\ref{tab:XGBoost}.

The class wise precision and recall for the random forest model are noted in Table~\ref{tab:random_forest_precision_recall}. F1-score for this model is 0.89. We report the confusion matrix in Table~\ref{tab:RandomForest}. The random forest model outperforms the XGBoost model.

\if{0}
\subsubsection{XGBoost Model}
\label{subsubsec:XGBoost_des}
 In this model, we use author profile based feature ($AP_{f}$), peer review based feature ($PF_{f}$) and network based features ($NE_{f}$) as model features. Specifically, we use $NE_{f}$ based features of three different network in our model. We found a few important features among all the features such as experience ($E_{cnt}$), degree centrality of $CCN$, H-index ($H_{ind}$), team size ($TS$), sentiment of review text ($SNT_{r}$), reviewer diversity ($R_{div}$), editor diversity ($Ed_{div}$), citation count, reciprocity of $CCN$, degree centrality of $CRN$, core, page rank of $CRN$. Confusion matrix is referrred to  
\fi

\begin{table}[tbhp]
\centering
\caption{Class wise precision and recall of the XGBoost model.}\label{tab:XGBoost_precision_recall}
\begin{adjustbox}{width=0.35\textwidth}
\begin{tabular}{|c|c|c|} \hline
{\bf Categories}&{\bf Precision} & {\bf Recall} \\ \hline
{\bf $ACC_{high}$} & 0.78 & 0.75  \\ \hline
{\bf $ACC_{mid}$} & 0.84 & 0.88  \\ \hline
{\bf $ACC_{low}$} & 0.92 & 0.88 \\ \hline 
\end{tabular}
\end{adjustbox}
\end{table}

\begin{table}[tbhp]
\centering
\caption{Confusion matrix of the XGBoost model.}\label{tab:XGBoost}
\begin{adjustbox}{width=0.35\textwidth}
\begin{tabular}{|c|c|c|c|} \hline
{\bf Categories}&{\bf $ACC_{high}$} & {\bf $ACC_{mid}$} & {\bf $ACC_{low}$}\\ \hline
{\bf $ACC_{high}$} & 2622 & 718 & 174 \\ \hline
{\bf $ACC_{mid}$} & 715 & 8669 & 502 \\ \hline
{\bf $ACC_{low}$} & 18 & 987 & 7389\\ \hline 
\end{tabular}
\end{adjustbox}
\end{table}

\if{0}
\subsubsection{Random Forest Model}
\label{subsubsec:random_forest_des}
In this model, we consider all the features for our model. We found random forest performs better over XGBoost algorithm. Out of all the feature, we found top important features. These are degree centrality of $CCN$, sentiment of review text ($SNT_{r}$), page rank of $CCN$, citation count, team size ($TS$), degree centrality of $CRN$, core, page rank of $CRN$, reciprocity of $CCN$, experience, H-index ($H_{ind}$), closeness centrality of $CCN$, betweenness centrality of $CCN$, 
 reviewer diversity ($R_{div}$). Confusion matrix of the model is givne in Table~\ref{tab:RandomForest}. We notice that random forest performs better than XGBoost classifier. Classwise precision and recall is reported in Table~\ref{tab:random_forest_precision_recall}. F1 score of this model is 0.89. F1 score of this model is relatively higher than F1 score of XGBoost model.
\fi

\begin{table}[tbhp]
\centering
\caption{Class wise precision and recall of the random forest model.}\label{tab:random_forest_precision_recall}
\begin{adjustbox}{width=0.35\textwidth}
\begin{tabular}{|c|c|c|} \hline
{\bf Categories}&{\bf Precision} & {\bf Recall} \\ \hline
{\bf $ACC_{high}$} & 0.82 & 0.82  \\ \hline
{\bf $ACC_{mid}$} & 0.87 & 0.91  \\ \hline
{\bf $ACC_{low}$} & 0.95 & 0.91 \\ \hline 
\end{tabular}
\end{adjustbox}
\end{table}
\vspace{-0.3cm}

\begin{table}[tbhp]
\centering
\caption{Confusion matrix of the random forest model.}\label{tab:RandomForest}
\begin{adjustbox}{width=0.35\textwidth}
\begin{tabular}{|c|c|c|c|} \hline
{\bf Categories}&{\bf $ACC_{high}$} & {\bf $ACC_{mid}$} & {\bf $ACC_{low}$}\\ \hline
{\bf $ACC_{high}$} & 2889 & 527 & 98 \\ \hline
{\bf $ACC_{mid}$} & 604 & 9006  & 276 \\ \hline
{\bf $ACC_{low}$} & 20 & 727 & 7647\\ \hline 
\end{tabular}
\end{adjustbox}
\end{table}

\begin{figure}[!tbh]
\centering
\setlength\tabcolsep{0pt}
\renewcommand{\arraystretch}{0.8}%
\begin{tabular}{@{}c@{}c@{}c@{}c@{}}
  \includegraphics[width=1\hsize]{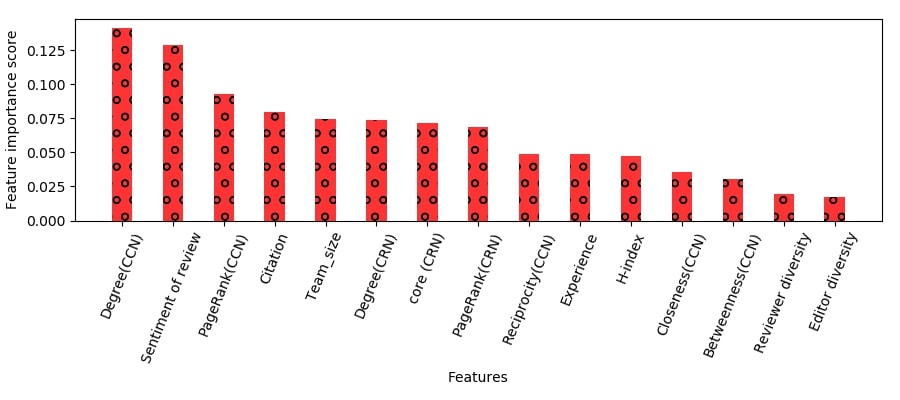} \\
\end{tabular}
\caption{Important features for the random forest model.}  
\label{fig:random_forest_feature_imp}
\end{figure}

\begin{figure}[!tbh]
\centering
\setlength\tabcolsep{0pt}
\renewcommand{\arraystretch}{0.8}%
\begin{tabular}{@{}c@{}c@{}c@{}c@{}}
  \includegraphics[width=1\hsize]{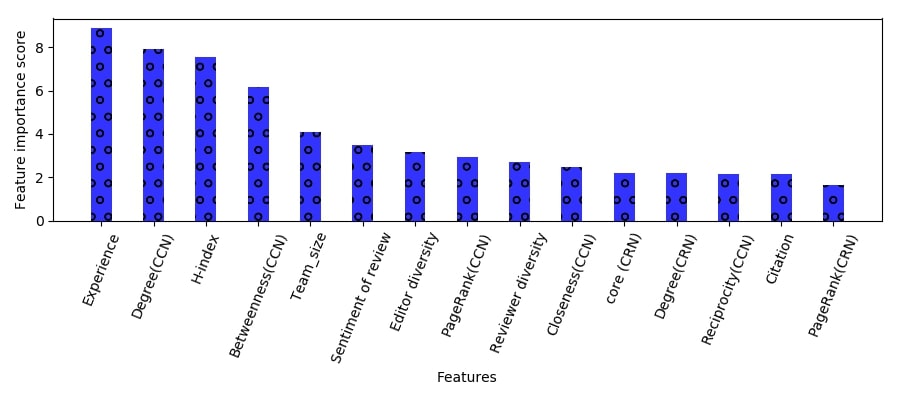} \\
\end{tabular}
\caption{Important features for XGBoost model.}  
\label{fig:xgboost_feature_imp}
\end{figure}

\noindent{\bf Feature importance}: Some of the important features for both the models are degree centrality of $CCN$, sentiment of review text ($SNT_{r}$), PageRank of $CCN$, citation count, team size ($TS$), degree centrality of $CRN$, core number, PageRank of $CRN$, reciprocity of $CCN$, experience, $h$-index ($H_{ind}$), closeness centrality of $CCN$, betweenness centrality of $CCN$, 
reviewer diversity ($R_{div}$). The individual set of features that are important for the two models are noted in Figure~\ref{fig:random_forest_feature_imp} (random forest) and Figure~\ref{fig:xgboost_feature_imp} (XGBoost).

\vspace{-0.3cm}
\vspace{-0.2cm}

\section{The role of the peer review system}
\label{sec:catwise_authors_editor_reviewer_analysis}
So far we have investigated author characteristics that could act as early indicators of the acceptance rate of the authors. However, recall, the reviewer and editor diversity measures presented in sections~\ref{ref_div} and~\ref{ed_div} respectively. In fact these features are also found to have strong predictive power in section~\ref{prediction}. Although we have used these features in profiling the authors, it can be easily reasoned that they are based on the functioning of the peer review system itself. In this section we shall therefore discuss the role of the peer review system (in any) in reinforcing the distinction among the three categories of authors. 

\begin{figure}[!tbh]
\centering
\setlength\tabcolsep{0pt}
\renewcommand{\arraystretch}{0.8}%
\begin{tabular}{@{}c@{}c@{}c@{}c@{}}
  \includegraphics[width=8cm, height=7cm,left]{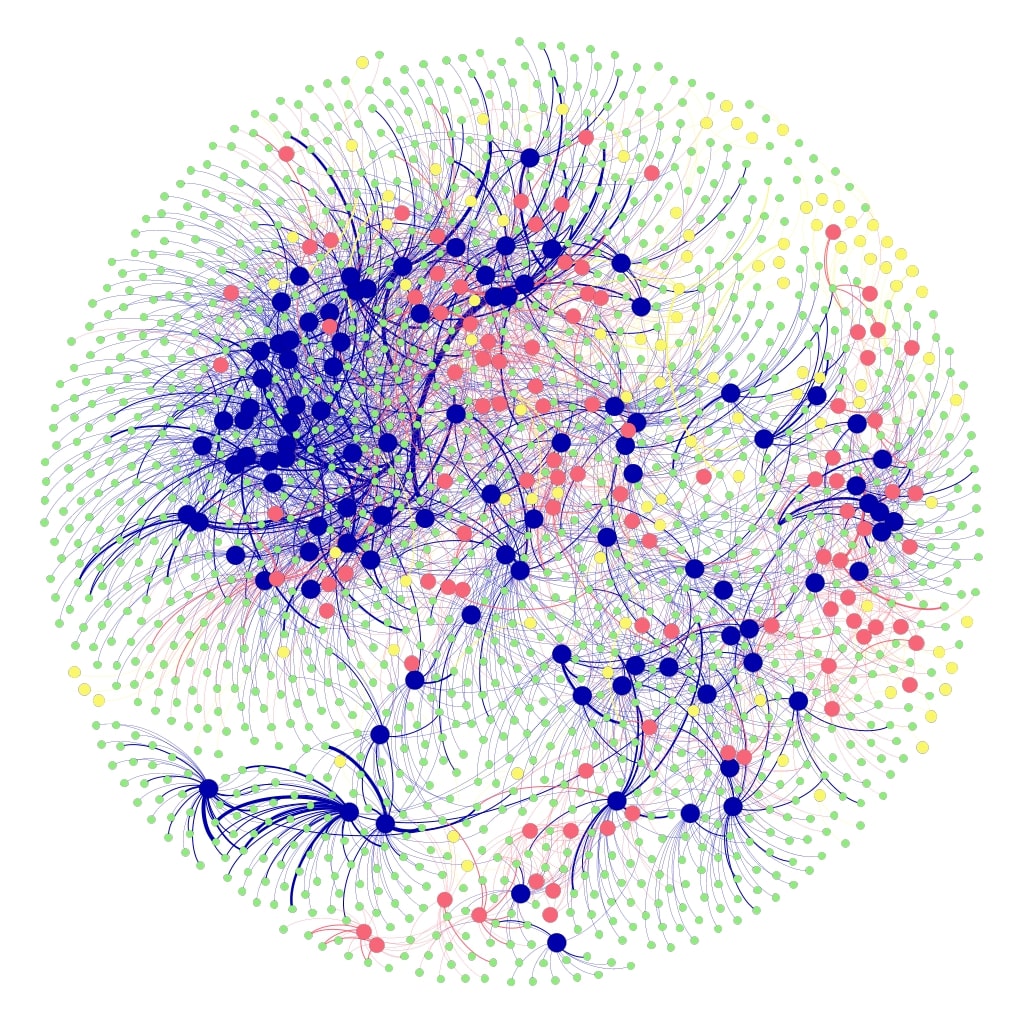}\\
  \includegraphics[width=6cm, height=0.45cm]{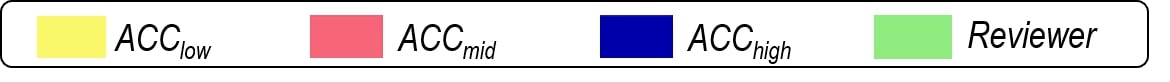}\\
  \includegraphics[width=8cm, height=8cm,left]{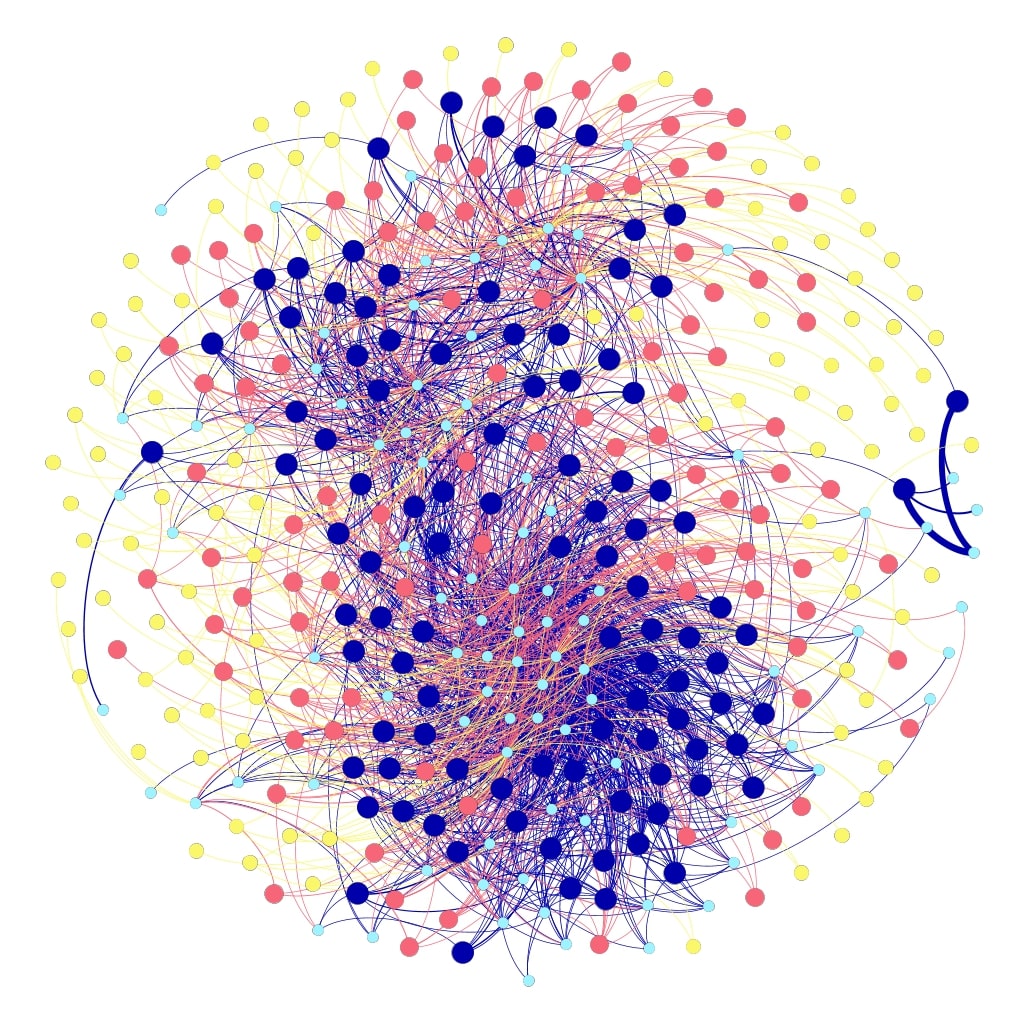}\\
 \includegraphics[width=6cm, height=0.3cm]{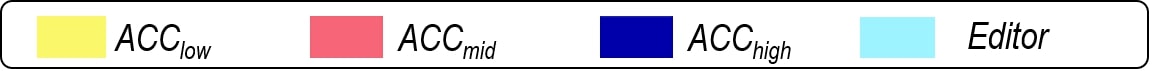}\\
\end{tabular}
\caption{(Up) This network shows the relationship between hundred top cited authors and their reviewer from three different categories. There is a directed edge from a reviewer to an author if the reviewer had reviewed one or more papers of the author. (Down) This network shows the relationship between hundred top cited authors and their editors from three different categories. There is a directed edge from an editor to an author if the editor had edited one or more papers of the author.}  
\label{fig:top_cited_hundred_authors}
\end{figure}

To this purpose, we characterize the authors of different categories in terms of the set of editors and reviewers who have ever edited/reviewed their paper. We consider pairs of authors from each category and compute the Jaccard overlap ($J$) of the reviewer and the editor sets respectively. Next for each category, we calculate the average pairwise $J$ values. Interestingly, for the reviewer set we observe that the average value of $J$ for $ACC_{high}$ authors is relatively higher (0.0202) compared to $ACC_{mid}$ (0.0016) and $ACC_{low}$ (0.0008) authors. For the editor set, the average value of $J$ for $ACC_{high}$ is 0.0302 whereas the average value for $ACC_{mid}$ and $ACC_{low}$ are similar (0.0137 and 0.0105 respectively). This potentially again indicates that there is less diversity in the editors and reviewers who are assigned to the $ACC_{high}$ category. However, one might argue that this could as well be an artefact of the authors in the $ACC_{high}$ category collaborating more heavily among themselves compared to the other two categories and therefore it is obvious that they would tend to have more overlap in the reviewer and editor sets. In order to verify if this is actually an artefact, we next consider for each category the pairs of authors who have never collaborated (i.e., never co-authored a paper together). For such pairs of authors in a category, we calculate the $J$ of their editor and reviewer sets again. In particular, we identify the \% of author pairs having $J$ in the range $[0.6, 1]$ and author pairs having $J$ exactly 1. We note the percentage overlap values in Table~\ref{tab:jaccard_overlap_editor_reviewer}. For both the editor and the reviewer sets we observe that even if the authors have never collaborated they tend to get more similar referees and editors in the $ACC_{high}$ category compared to the other two categories. This result indicates that the initial observation that we made was not an artefact and that the peer-review system indeed enables a less diverse referee and editor set for the $ACC_{high}$ authors. We present a visualisation of this phenomenon in Figure~\ref{fig:top_cited_hundred_authors}. In  Figure~\ref{fig:top_cited_hundred_authors} (Up), the green coloured nodes represent the reviewers and the blue, the red and the yellow nodes correspond to the authors in the $ACC_{high}$, $ACC_{mid}$ and $ACC_{low}$ categories respectively. There is a directed edge from a reviewer to an author if the reviewer had reviewed one or more papers of the author (i.e., a directed bipartite network). The visualisation again indicates that there are `patches' of clusters of unique reviewers around authors of the $ACC_{high}$ category. Similarly, in Figure~\ref{fig:top_cited_hundred_authors} (Down) the sky blue colored nodes represent the editors and the blue, the red and the yellow nodes correspond to the authors in the $ACC_{high}$, $ACC_{mid}$ and $ACC_{low}$ categories respectively. There is a directed edge from an editor to an author if the editor had edited one or more papers of the author. Similar patches of clusters also appear here. Overall, we believe that this might lead to potential discrimination and unfairness and should therefore be further investigated by the system admins.

\begin{table}[tbhp]
\centering
\caption{Percentage of author pairs having Jaccard overlap of editor and reviewer set in $[0.6, 1]$ and exactly 1. }\label{tab:jaccard_overlap_editor_reviewer}
\begin{adjustbox}{width=0.45\textwidth}
\begin{tabular}{|c|c|c|c|c|} \hline
{\bf Categories}&\multicolumn{2}{c|}{\bf Editor set} & \multicolumn{2}{c|}{\bf Reviewer set} \\ \hline
{} & $J$ ($[0.6,1]$) & $J$ ($= 1$) & $J$ ($[0.6,1]$) & $J$ ($= 1$) \\ \hline
{\bf $ACC_{high}$} &  1.94\%  & 1.86\% & 0.46\% &0.31\%\\ \hline
{\bf $ACC_{mid}$} & 1.36\%  & 1.31\% & 0.32\% &  0.13\%\\ \hline
{\bf $ACC_{low}$} & 1.03\% & 1.02\% &  0.04\% & 0.03\%\\ \hline 
\end{tabular}
\end{adjustbox}
\end{table}


 As an additional investigation we choose author pairs across categories and observe how their editor sets overlap. If we choose author pairs with one from $ACC_{high}$ and another from $ACC_{mid}$ the $J$ value in $[0.6, 1]$ for the editor set is 0.13\%. Similarly, if we choose author pairs with one from $ACC_{mid}$ and another from $ACC_{low}$ the $J$ value in $[0.6, 1]$ for the editor set is 0.57\%. However, what is most intriguing is that if we choose author pairs with one from $ACC_{high}$ and another from $ACC_{low}$ the $J$ value in $[0.6, 1]$ for the editor set is 0\%. This indicates that the editors who are assigned to the $ACC_{high}$ category of authors are almost never assigned to the $ACC_{low}$ category users. Once again this could be indicative of a potential unfairness situation in the peer-review system and needs to be carefully investigated further.


\vspace{-0.2cm}
\section{Related Work}
\label{sec:related_work}
Peer review system plays an important role in the acceptance of a research paper in a journal. Quality peer review system helps authors to improve themselves. There are lots of debates on the quality~\cite{Jefferson:2002} and bias\footnote{https://www.nature.com/news/let-s-make-peer-review-scientific-1.20194} in a peer review system~\cite{Huisman:2017,Falkenberg:2018,Sikdar:2016}. Jefferson {et al.}~\cite{Jefferson:2002} investigated the quality of editorial peer review. They claimed that measuring the quality of peer review require huge co-operation of authors. Sikdar {et al.}~\cite{Sikdar:2017} studied reviewer-reviewer interaction network to predict the long term citation of a paper. They also studied whether the peer review system can be improved.  In~\cite{Sikdar:2016}, the authors investigated anomalies in a peer review system. They computed different features from the editor and the reviewer information available. In~\cite{helmer17} the authors investigated the existence of gender bias in a peer review system. Another interesting study by Tomkins {et al.}~\cite{tomkins17} showed that a single blind reviewing system gives disproportionate advantage to the papers of famous authors and authors from highly reputed institutions. In similar lines the authors in~\cite{alina19} proposed how to improve a single blind review process.  

Earlier research also explored various author profile based features such as experience, citation count, $h$-index, research topic diversity to quantify research productivity/success of an author~\cite{Bu:2018}. The productivity of an author~\cite{Abramo:2018} had been defined as the extent of his/her contribution (publications) to the scientific community. Most of the earlier research focused on whether such author profile based features are sufficient to justify ones research productivity. In~\cite{Bremholm:2005}, the authors explored the productivity of authors and their citations considering publications in the Proceedings of the Oklahoma Academy of Science (POAS). They found that authors with high productivity are not highly cited. Bayer {et al.}~\cite{Alan:1966} computed citation count to measure the productivity and found that it is less correlated with the quality of researcher's academic career but there is no correlation with his/her IQ. 

Our work is very different from the above studies. We utilise author profile information, peer review information and three different networks to predict the class of an author based on his/her acceptance rate. 




\vspace{-0.3cm}
\section{Conclusion}
\label{conc}
We categorize the authors into three classes based on their acceptance rate in the journal. We characterise these classes of authors based on their profile, the peer reviews their papers received and three different networks. The authors with high acceptance rate seem to be markedly different in terms of many of these characteristic features. Finally, using these features we show that it is possible to predict the acceptance rate class early for any author. 

In future we would like to investigate in more details the reasons for the differences in the reviewer and editor diversities across the classes. In specific this problem can be posed as an anomaly/bias detection where we plan to use state-of-the-art techniques to understand the precise reasons for such uneven diversity across the classes.
\vspace{-0.25cm}
\section{Acknowledgements}
We thank Media Lab SISSA for providing us with the necessary JHEP data for the analysis. RH and AM thank Simons Foundation for financial support through the Simons Associateship Programme.

\vspace{-0.25cm}
\bibliographystyle{ACM-Reference-Format}
\bibliography{ref}

\end{document}